# Data and Clock Transmission Interface for the WCDA in LHAASO

**S.P. Chu,[a,b] L. Zhao,[a,b,*] Z.Y. Jiang,[a,b] C. Ma,[a,b] X.S. Gao,[a,b] Y.F. Yang,[a,b] S.B. Liu[a,b] and Q. An[a,b]**

[a] *State Key Laboratory of Particle Detection and Electronics, University of Science and Technology of China, Hefei City, Anhui Province, CHINA*

[b] *Department of Modern Physics, University of Science and Technology of China, Hefei City, Anhui Province, CHINA*

*E-mail*: zlei@ustc.edu.cn

ABSTRACT: The Water Cherenkov Detector Array (WCDA) is one of the major components of the Large High Altitude Air Shower Observatory (LHAASO). In the WCDA, 3600 Photomultiplier Tubes (PMTs) and the Front End Electronics (FEEs) are scattered over a 90000 $m^2$ area, while high precision time measurements (0.5 ns RMS) are required in the readout electronics. To meet this requirement, the clock has to be distributed to the FEEs with high precision. Due to the "triggerless" architecture, high speed data transfer is required based on the TCP/IP protocol. To simplify the readout electronics architecture and be consistent with the whole LHAASO readout electronics, the White Rabbit (WR) switches are used to transfer clock, data, and commands via a single fiber of about 400 meters. In this paper, a prototype of data and clock transmission interface for LHAASO WCDA is developed. The performance tests are conducted and the results indicate that the clock synchronization precision of the data and clock transmission is better than 50 ps. The data transmission throughput can reach 400 Mbps for one FEE board and 180 Mbps for 4 FEE boards sharing one up link port in WR switch, which is better than the requirement of the LHAASO WCDA.

* Corresponding author.

## Contents

## 1. Introduction

The Large High Altitude Air Shower Observatory (LHAASO) project is a large air shower particle detector array aiming for gamma ray source surveys above 100 GeV [1]. As shown in figure 1, it consists of a 1-km$^2$ Extensive Air Shower Array (KM2A), a Water Cherenkov Detector Array (WCDA), a Shower Core Detector Array (SCDA), and a Wide Field of view Cherenkov Telescope Array (WFCTA) [2]. As one of the major components in LHAASO, the WCDA [3] consists of four 150 m × 150 m water ponds, and 3600 Photomultiplier Tubes (PMTs) are placed under water at 4 m depth. The PMT output signal would vary from 1 Photo Electron (P.E.) to 4000 P.E., and both high precision charge and time measurements are required over this range. The required charge resolution is better than 30% @ 1P.E. and 3% @ 4000 P.E., while the time resolution is required to be better than 500 ps RMS over the whole range [4].

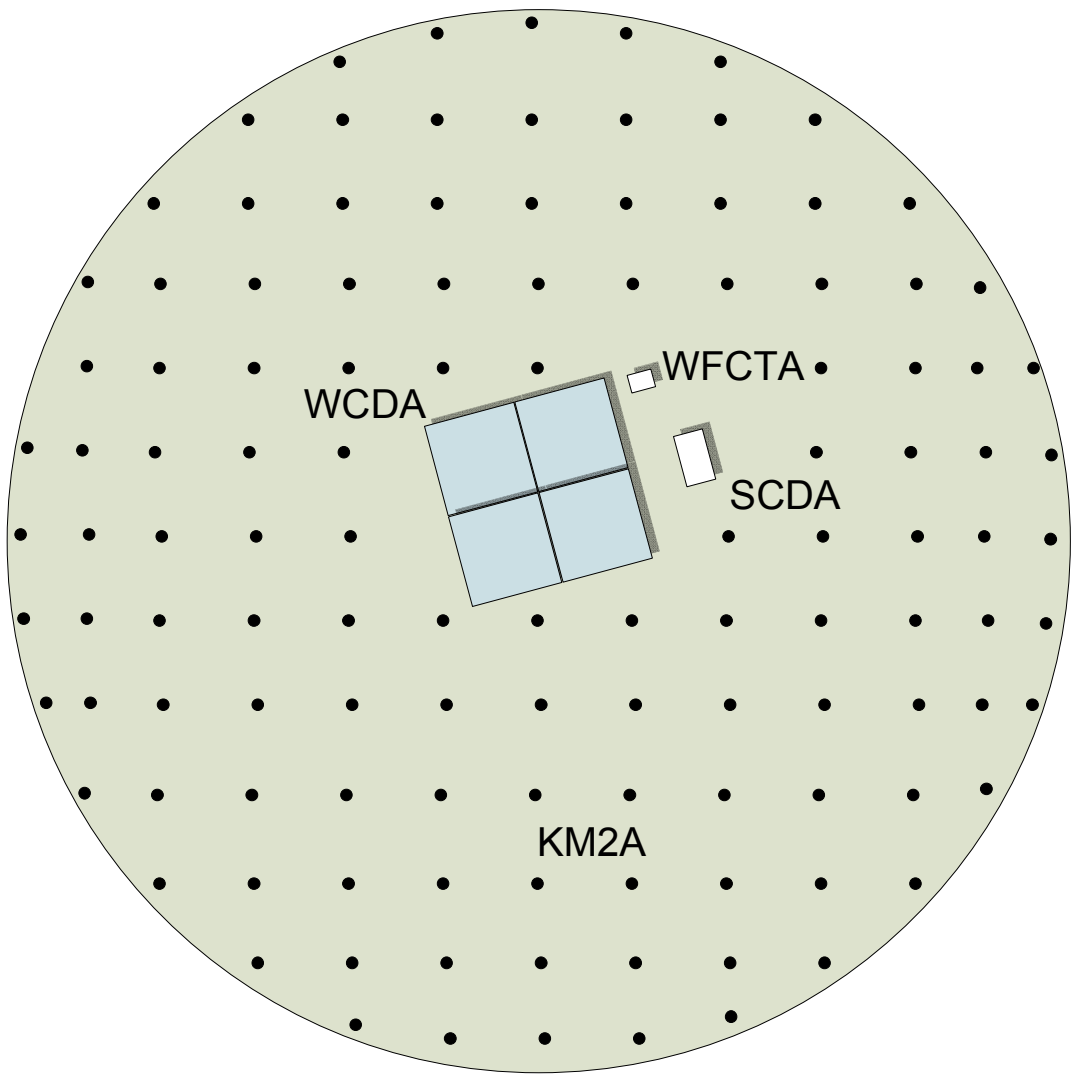

**Figure 1.** Layout of the LHAASO complexity.

Considering that 3600 PMTs are scattered over a large area of 90000 $m^2$, to avoid signal attenuation and Signal-to-Noise Ratio (SNR) deterioration caused by signal transmission over long cables, a distributed architecture is proposed for the readout electronics of LHAASO WCDA: 400 FEEs are placed just above water near PMTs, and signal processing and digitization are integrated in these FEEs. This means that clock and data need to be transferred over long distance, and hence come challenges.

First, a high time resolution better than 0.5 ns is required for the whole electronics of WCDA, which means that a high quality clock signal needs to be distributed to the FEE nodes, automatic clock phase alignment is required with varying ambient temperatures, and the compensation precision is aimed to be better than 100 ps. On the other hand, due to the "triggerless" architecture [5, 6], all the raw data on the FEE are required to be transferred to Data Acquisition (DAQ), thus high speed data transfer interface based on TCP/IP standard is necessary. Meanwhile, to simplify the architecture of the readout electronics, mixed transmission of clock, data, and commands based on the same fiber is preferred.

In LHAASO, the White Rabbit (WR) [7] technique is adopted for clock and data transmission of all the electronics of different detectors. As shown in figure 2, the WR switches are used to transmit the clock signal, and meanwhile collect the data from the FEEs. However, the WR technique features clock compensation with a precision of around 0.5 ns [7], which is unable to meet the requirement of the WCDA electronics. Another problem is that no functionality is provided for high speed data packaging based on TCP/IP protocol in the WR method. To solve the above problems, we devote efforts on the clock and data transmission interface design, and make it compatible with the WR switches in the system.

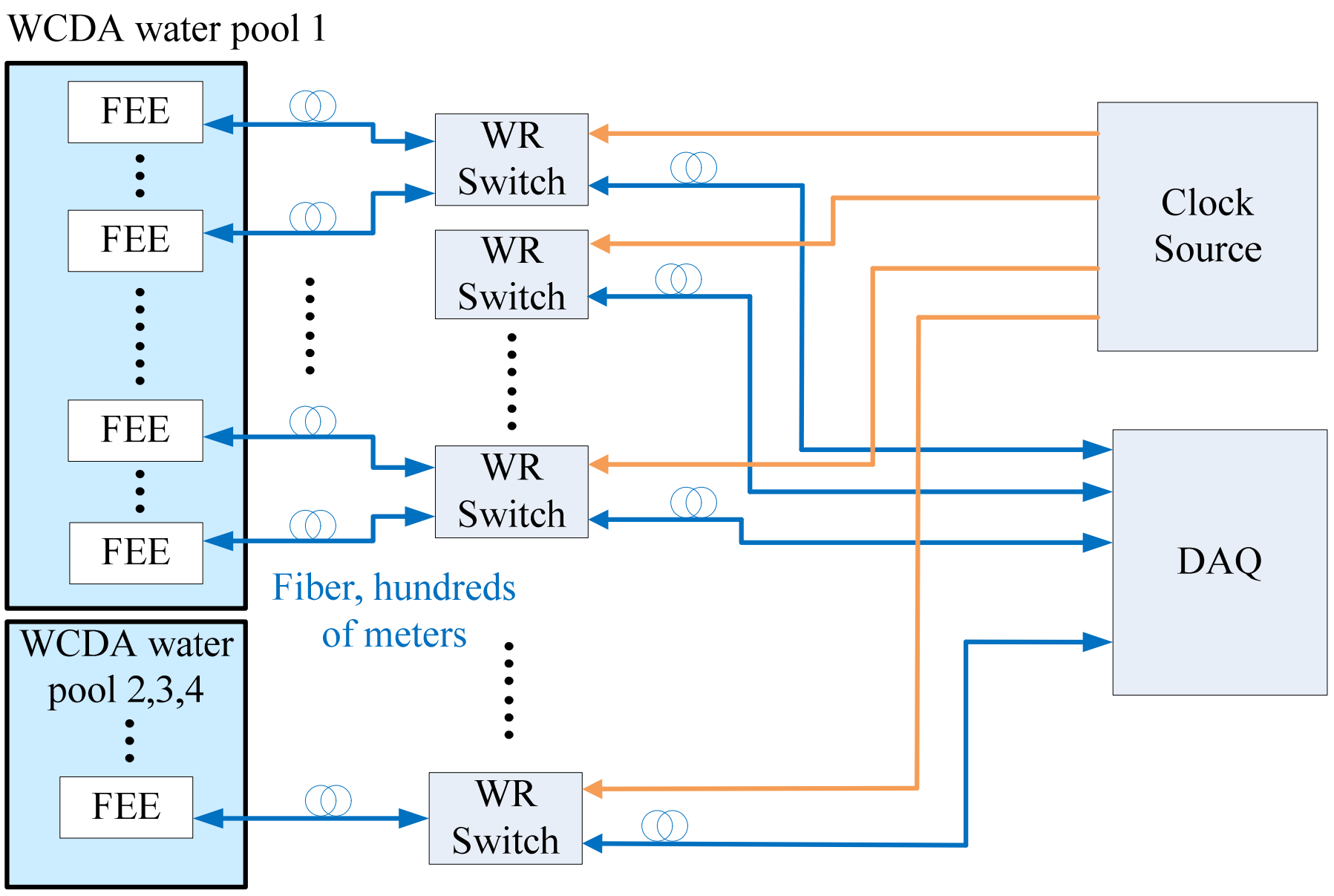

**Figure 2.** Architecture of LHAASO WCDA readout electronics

This paper is organized as follows. In Section II, we present the structure of the data and clock transmission interface. In section III, we conducted tests to evaluate its performance. And this paper is concluded in section IV.

## 2. Architecture

Figure 3 shows the structure of the data and clock transmission interface. As mentioned above, to simplify the system architecture, the commands, data and clock are transmitted via one single-mode fiber, and different light wavelengths are used in two directions (upwards: from FEE to WR switch, downwards: from the WR switch to the FEE). To simplify the electronics, we use the GTP [8] interface in FPGA (Field Programmable Gate Array) instead of external SerDes (Serializer-Deserializer) chip to mix the data and clock signal within one serial data stream. In the upwards direction, the data (i.e. time and charge measurement results) on the FEE and clock are imported to GTP and further sent to WR switches through fibers, while in the downwards direction, the GTP recovers the clock and data from the serial streams coming from WR switches. As for the data streams, there are two different types of data packets. One is the normal data packet based on the TCP/IP, which is either the data of measurement results from FEE to DAQ or commands from DAQ. The other is WR PTP packet, which contains the information used for clock synchronization.

To achieve clock phase compensation, the FEE transmits the recovered clock back to the WR switch, and the roundtrip delay of the clock signal between WR switch and FEE can be measured. This delay information is just contained in the WR PTP packets. The FEE receives these packets and adjusts the clock phase using “phase adjustment” for compensation. As mentioned above, to achieve a compensation precision beyond the WR technique, we proposed a new method based on delay increment compensation [9]. However, our work in [9] focused on the algorithm of this method, and only simple prototypes were designed to verify the basic idea. To apply this method in the WCDA of LHAASO, in our recent work, we improved and extended the hardware design of clock interface, and more efforts were devoted on high-speed data transfer and integrating the data and clock transmission together in one interface module, as well as making it compatible with the WR switches. To be compatible with the WR protocol, the WRPC (White Rabbit PTP Core) [10] is required to be integrated, which also distributes these two types of data packets. In order to guarantee data integrity in transmission, we use SiTCP [11] processor for data packaging based on the TCP/IP protocol. To construct a complete data and clock transmission interface, special bridges were designed, e.g. "bridge logic" between SiTCP and WRPC in figure 3.

Besides, considering that in actual application the 400 FEEs are scattered over 90000 m$^2$, the capability of remote logic update of the FPGA is also demanded.

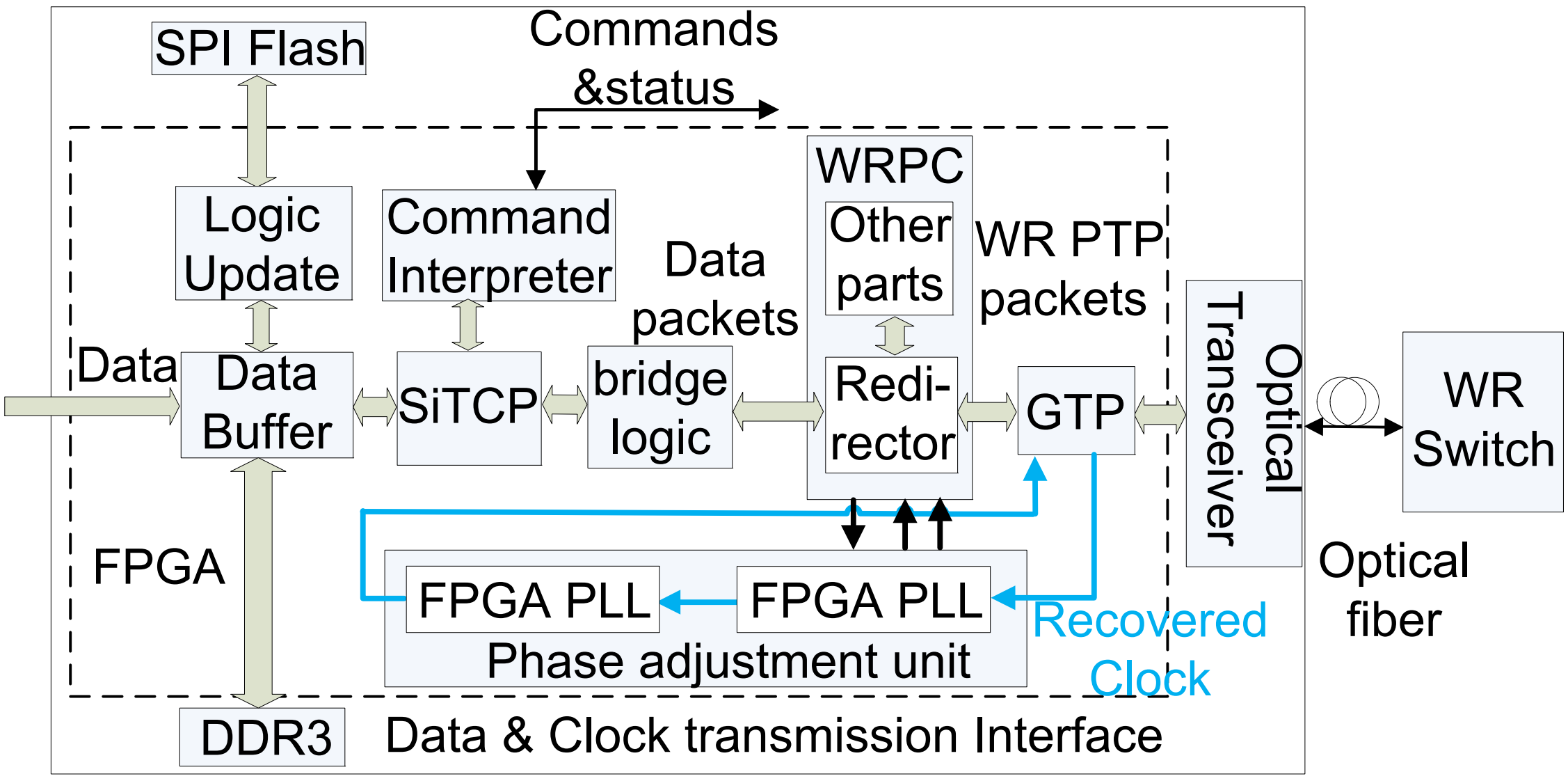

**Figure 3.** Structure of the data and clock transmission interface

### 2.1 Clock Synchronization

As shown in figure 4, the clock compensation module consists of two parts: one is the phase adjustment unit and the other is the WRPC that interfaces with the WR switch.

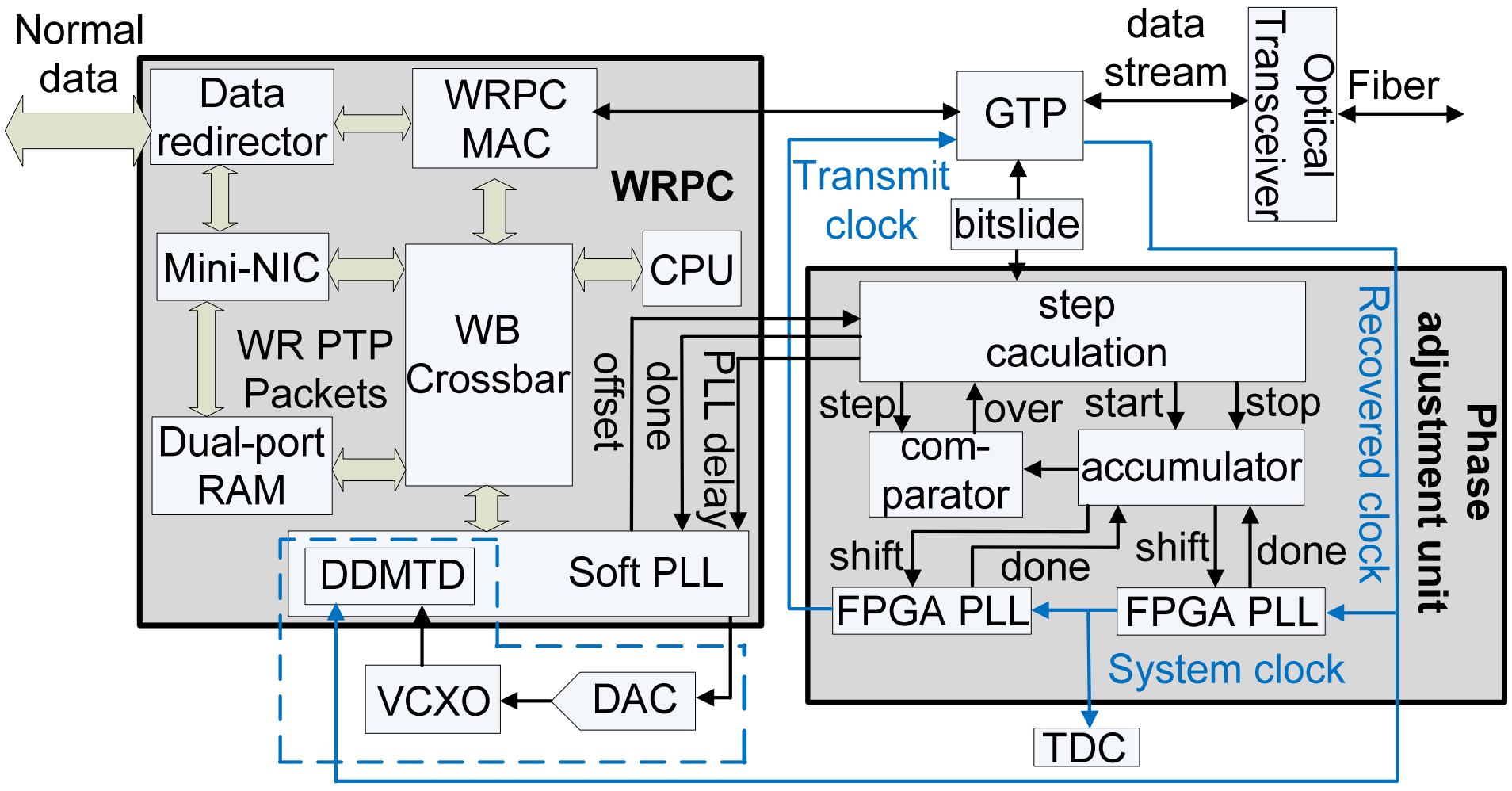

**Figure 4.** Structure of the clock compensation module

As shown in figure 4, the WRPC consists of several parts interconnected by the Wishbone Crossbar and managed by a CPU (LatticeMicro32, LM32). Figure 5 shows the structure of the WRPC software running on the CPU. And there are two main tasks for this software. One is to communicate with the WR switch by PTP packets, and obtain the information of certain time points [7] to calculate the roundtrip delay. As mentioned above, this roundtrip delay includes the upwards and downwards delays contributed by the fiber and the electronics. In the WR method, the delay of the electronics is considered as a constant value, and then the downwards delay caused by the fiber can be calculated using the fiber asymmetry coefficient [7], and this downwards delay just corresponds to the phase compensation value (i.e. "Delay Offset" in figure 5). The other task is to compensate the clock phase. The software outputs digital control codes for an external DAC, and further tunes the frequency and phase of the VCXO. To evaluate the compensation effect, the compensated clock (i.e. the output of the VCXO) and the recovered clock (i.e. the clock before compensation) are fed to a phase detector, which is actually implemented by the DDMTD (Digital Dual Mixer Time Difference) in the FPGA. The results of the DDMTD are transferred to the software via an interrupt operation. By comparing the information from DDMTD and the "Delay offset", the software adjusts the control codes of DAC until the expected compensation is completed. Therefore, this is actually a soft PLL structure.

Since the delay of the electronics is defined as a constant value, the WR method cannot deal with the situation in varying temperature. An improved method is to add thermometer and correct the temperature effect using Look-Up Tables (LUTs) [12], but this method is still too complex. As for the phase compensation circuits, it is also too complex, requiring external chips, e.g. DAC, VCXO, which limits the compensation effect.

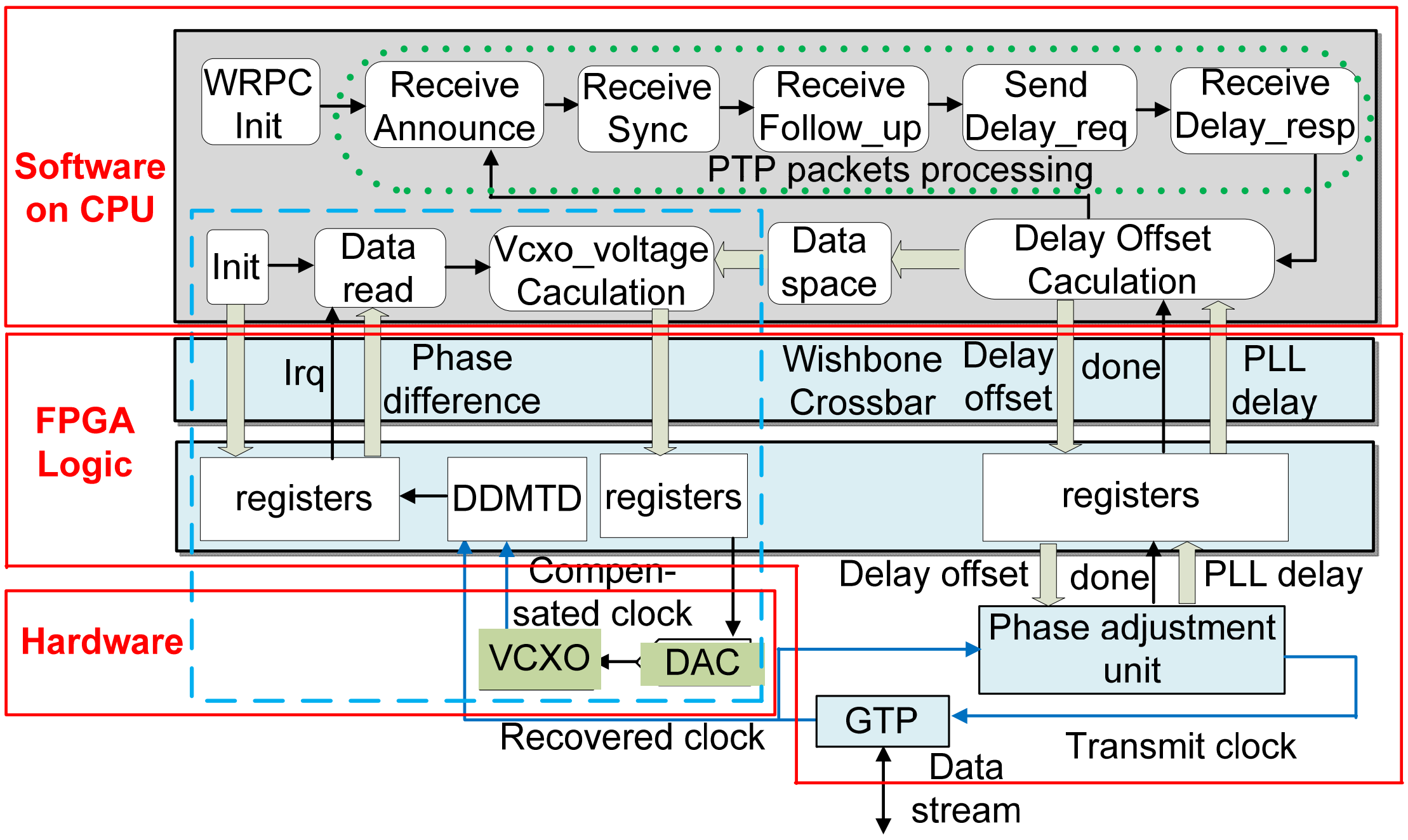

**Figure 5.** Structure of WRPC software running on the CPU

To overcome these obstacles, we proposed an enhance method for clock compensation [9]. With our method, the original software is revised to calculate the downwards and upwards delay increment values with a much simpler structure [9]. As for compensation circuits, we abandoned the complex soft PLL, removing the external DAC and VCXO, etc., and the DDMTD in the FPGA. Instead, we used two internal PLLs [13] in Artix-7 FPGA for phase compensation. The corresponding control logic is thus also simplified, as shown in figure 4. In placement and routing during FPGA logic design, special attention was paid on the positions of these two PLLs to implement an approximately symmetrical structure. The phase adjustment step size of the PLL is about 15 ps, and the "phase adjustment unit" in figure 4 is used to control the PLLs. Since the clock phase of the PLL can only be adjusted by one step each time, an accumulator is implemented to calculate whether the expected phase adjustment value has been reached.

### 2.2 Data and commands transmission

As aforementioned, due to the "triggerless" architecture, all the data on the FEE (charge and time measurement results) need to be transferred to the DAQ. To guarantee the data reliability, data packaging based on TCP/IP protocol is required. As shown in figure 6, the FEE data are first stored in the "Data buffer", and then sent to the SiTCP core for data packaging. To deal with the situation of transient bursts of data rate caused by cosmic rays, external DDR3 DRAMs are employed as additional data cache. To be compatible with the WR protocol, an interface bridge between SiTCP and WRPC is designed. The output data from WRPC are then converted to a serial data stream by GTP and further sent to the WR switch through an optical fiber.

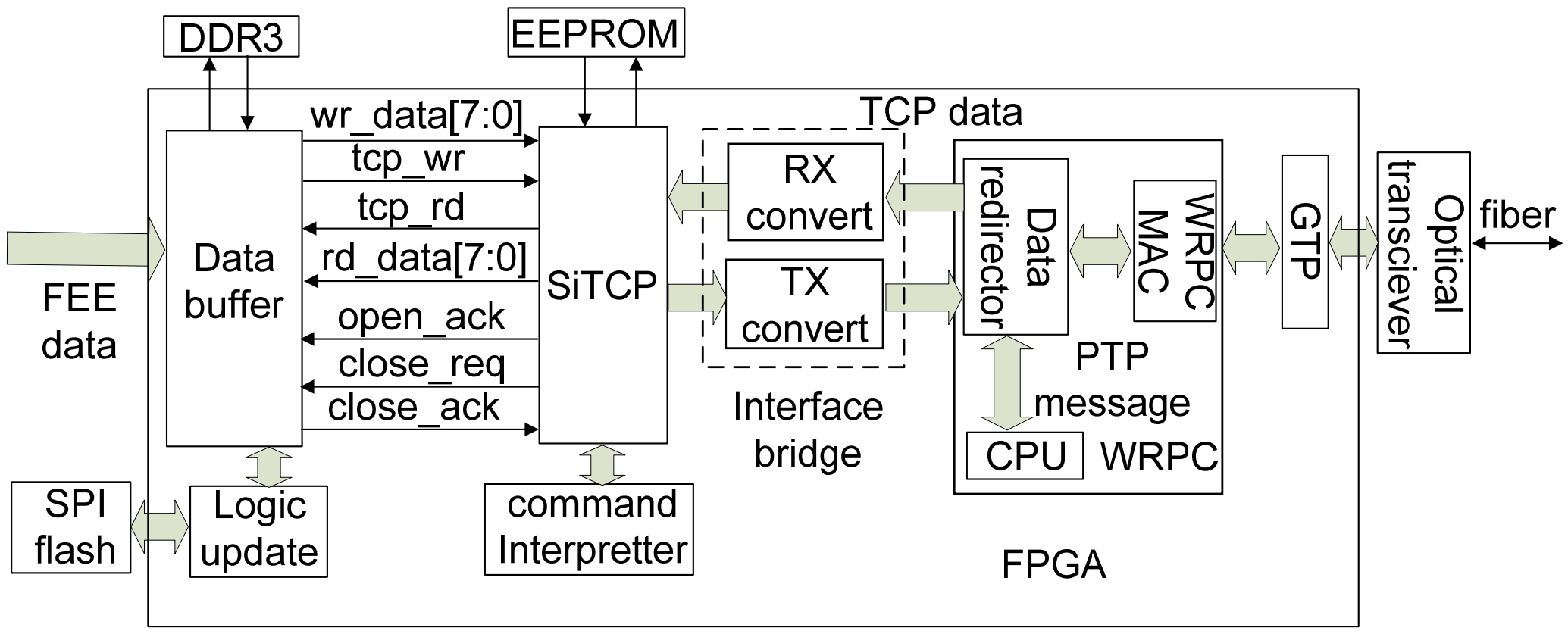

**Figure 6.** Structure of the data transmission interface

#### 2.2.1 Data buffering

Both charge and time measurement of the PMT output signals are required in FEE. To achieve a large dynamic range from 1 Photo Electron (P.E.) to 4000 P.E., two channels are used for each PMT: one is used to read out the signal from the anode, and the other is for a dynode. As for time measurement, two thresholds are employed in discrimination circuits [14]. To correlate with other cosmic ray observatories in the world, the time measurement results also need to contain the information with a bin size of 333 ps and a measurement range up to second, hour, month and year. All these data need to be transferred to DAQ, and we organize these data in a 128-bit data frame for each PMT channel. With an average hit rate of around 50 kHz, the corresponding data rate of each FEE is around 57.6 Mbps. As shown in figure 7, these measurement results for each PMT channel is first buffered by the corresponding FIFO, and then assembled with the other channels in the “Input FIFO”. As mentioned above, external DDR3 DRAMs (MT41J512M8 from Micron Technology, Inc.) is used as additional data cache.

Besides the data transmission, the two different types of messages from DAQ are received through the TCP and UDP interface of the SiTCP, respectively. One is the update logic configuration data of the FPGA, which is bulk data transferred through the TCP interface. The other is the command message received through UDP interface, which is further interpreted to different commands, as shown in figure 7.

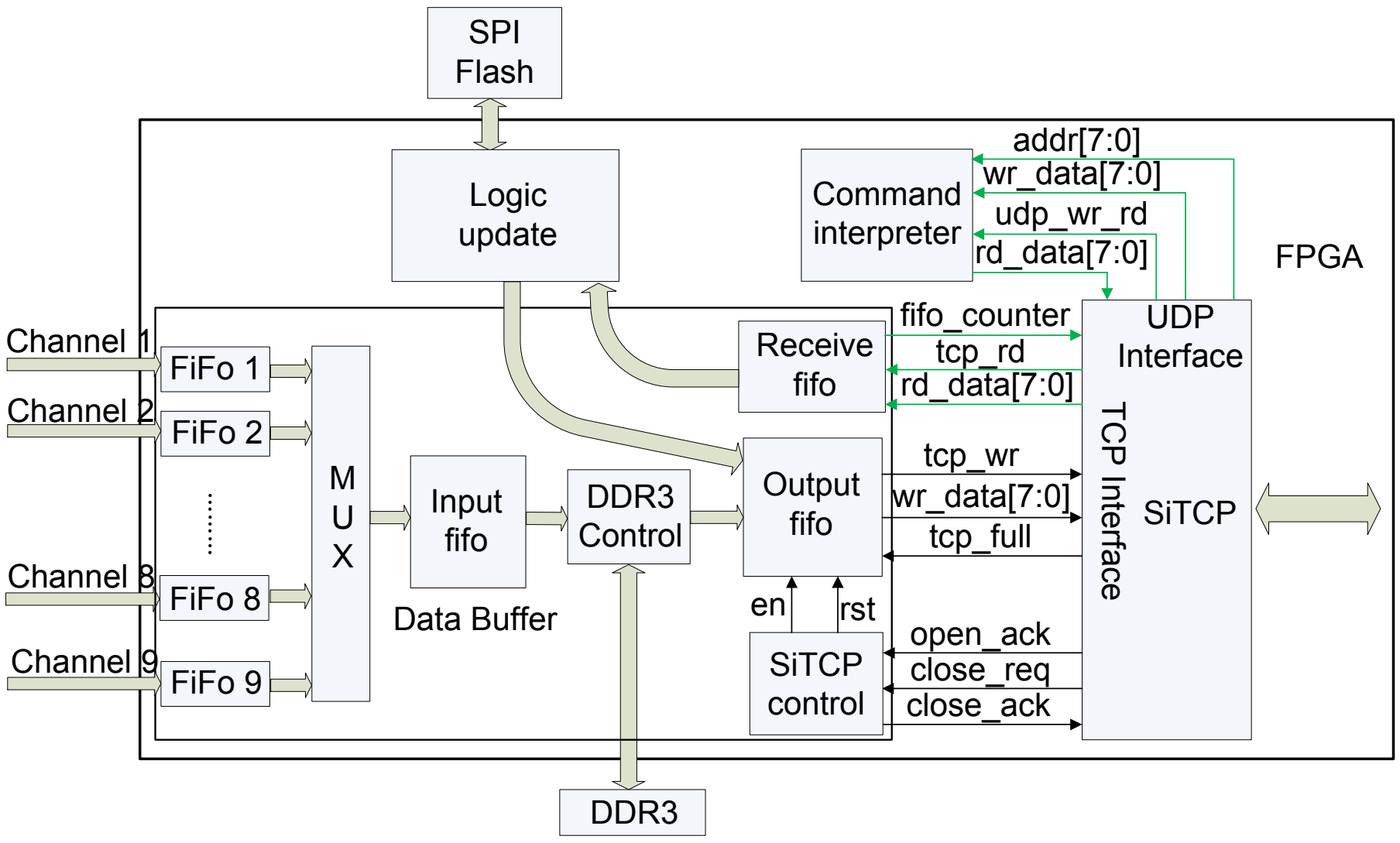

**Figure 7.** Data buffering and commands receiving

### 2.2.2 SiTCP and Interface Bridge

After buffering, the data are sent to the SiTCP, which is a hardware-based TCP processor optimized for the detector systems [11]. The standard TCP/IP communication protocol set is large and complex, and includes management protocols that are not necessary in readout electronics for physics experiments featuring much simpler network. By reducing the protocol set, the SiTCP has a small circuit size, making it easy to be implemented in a single FPGA device.

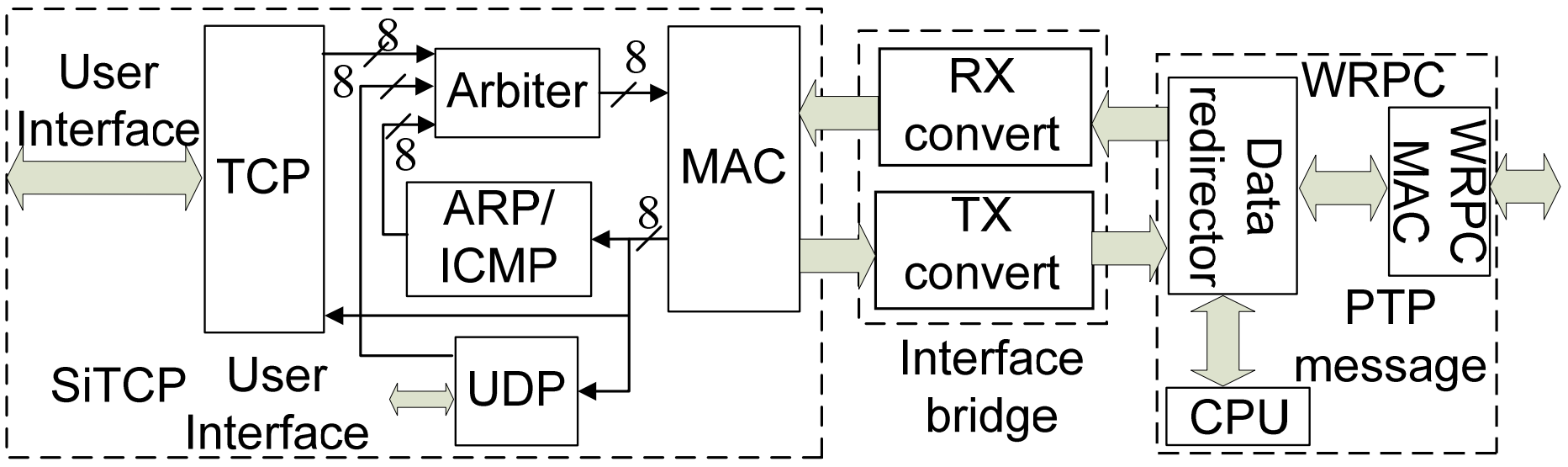

**Figure 8.** Block diagram of data packaging using SiTCP and WRPC

However, SiTCP itself is not compatible with WR protocol. As shown in figure 8, the SiTCP consists of MAC, TCP, ARP/ICMP, UPD and Arbiter blocks. To communicate with the WPRC, an interface bridge is necessary. The SiTCP packages the data, sends it to the WRPC in the “TX” direction, receives data from the WRPC and unpacks it in the “RX” direction. And a GMII (in SiTCP part) to/from Wishbone (in WRPC part) interface bridge logic is designed. Meanwhile, the WRPC and SiTCP both contain a MAC unit inside. This means that the data frame from the SiTCP has to be stripped of the header and tail in the Data Link layer of OSI (Open System Interconnection) before being transferred to WRPC, while in the opposite transfer direction, the frame from WRPC needs to be packaged again before input to SiTCP. This function is also integrated in the “interface bridge” block in figure 8.

The figure 9 shows the structure of the interface bridge logic. As for the “TX” direction, the frame generated by the SiTCP is first stripped of Preamble, SFD (Start of Frame Delimiter), Destination address, Source address, Length/Type and FCS (Frame Check Sequence). And then a FIFO is added to buffer the data to prevent data loss when the WRPC is transferring WR PTP message simultaneously. And then the data is converted and sent to the Wishbone interface of the WRPC. As for the “RX” direction, the processing method is similar except that a CRC (Cyclic Redundancy Check) generator is designed to add the FCS at the end of the frame.

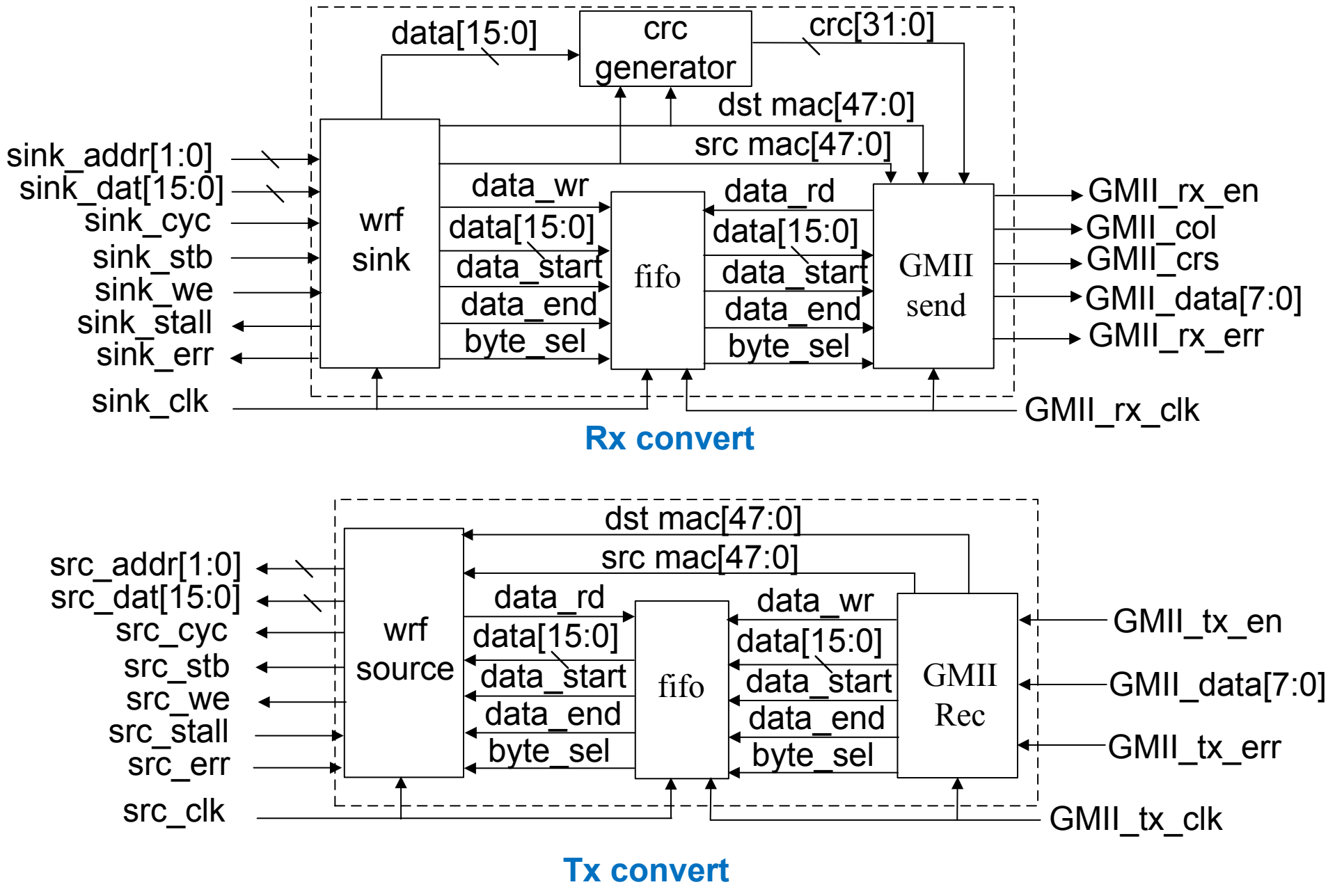

**Figure 9.** Structure of Interface Bridge.

### 2.2.3 Remote Logic Update

In the actual application, hundreds of FEEs are scattered over a 90000 m$^2$ area, and these FEEs are places in sealed metal box (four FEEs in each) above the water ponds of WCDA. Therefore, it is very difficult for manual updating of the FPGA logic in FEEs through JTAG (Joint Test Action Group) interfaces after system installation is finished. So the capability of remote logic update [15] through the fiber is important.

As shown in figure 10, the original logic bitstream (FPGA configuration data) is stored in a 256 Mb Flash (N25Q256A) with the SPI (Serial Peripheral Interface) interface, and is loaded automatically into the FPGA at power-on. When logic update is needed, the SiTCP will receive a special 256-Byte packet from DAQ, which is buffered in the “receive FIFO” and then fed to “unpackaging” block. The token information is extracted from the first four Bytes, and interpreted to different actions. The first command is to erase the content of the flash which stores the FPGA configuration data. To guarantee the robustness of the system, the space of this flash is divided into two parts, and the first part stores a verified stable logic, while the second part is for FPGA logic update. So the erasing operation is only valid for the second part in the flash. In the second step, continuous writing packets are received from the SiTCP, with which the second part of the flash is refreshed. To make sure the correctness of the writing operation, the content is also read back to the DAQ for data checking. If no errors are observed, a reset operation is started, which actually loads the updated logic into the FPGA device.

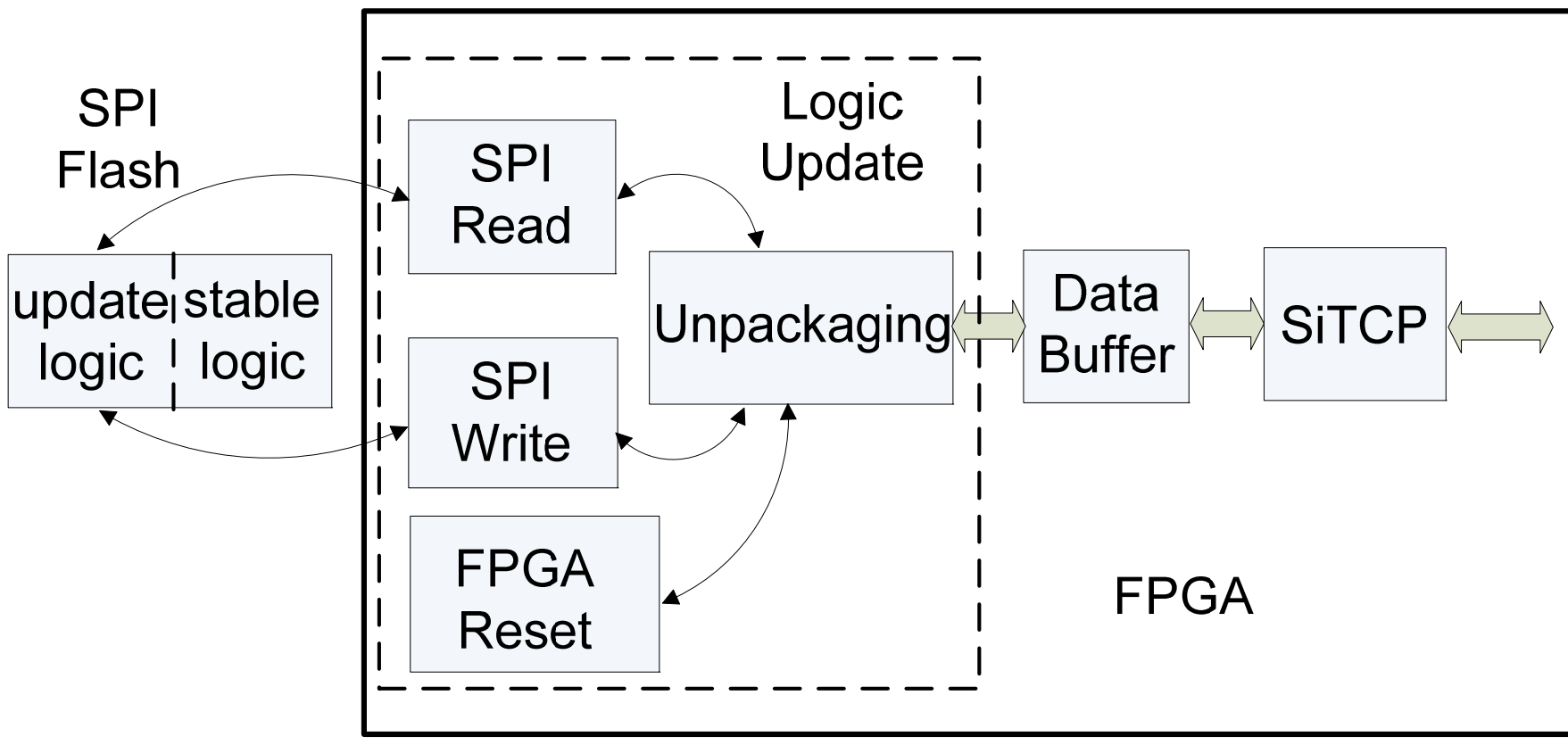

**Figure 10.** Structure of the remote Logic Update

## 3. Tests and Results

To evaluate the performance of the clock and data transfer interface, we designed a prototype and conducted a series of tests in the laboratory.

### 3.1 Performance of Clock transmission and Synchronization

As mentioned above, high precision clock synchronization is very important for time measurement in LHAASO WCDA. The clock synchronization performance is tested in two steps: synchronization test with multiple power-up operations and varying temperature.

In the first step, a straight forward idea is to power on the system in multiple times and observe whether the clock phase can be automatically synchronized. As shown in figure 11, We used a 400-meter fiber to connect our prototype with a WR switch, and observed the clock signals of the prototype and the switch using a high speed oscillator (104MXi from LeCory corporation), and the phase difference variation corresponds directly to the compensation effect.

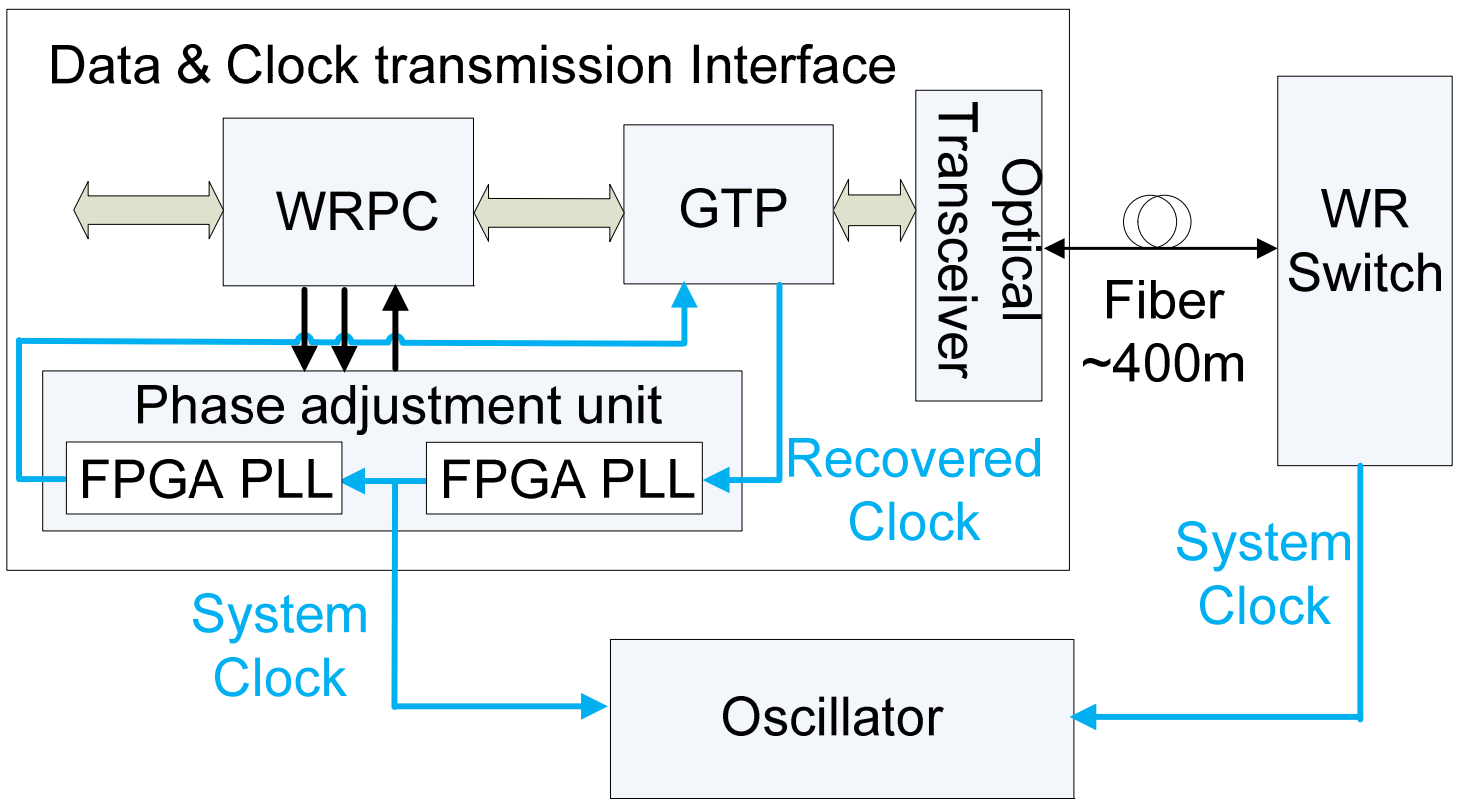

**Figure 11.** Test bench to evaluate the clock compensation performance.

As shown in figure 12, the clock phase difference between the data and clock transmission prototype and WR switch is very stable, with a maximum variation of 50 ps.

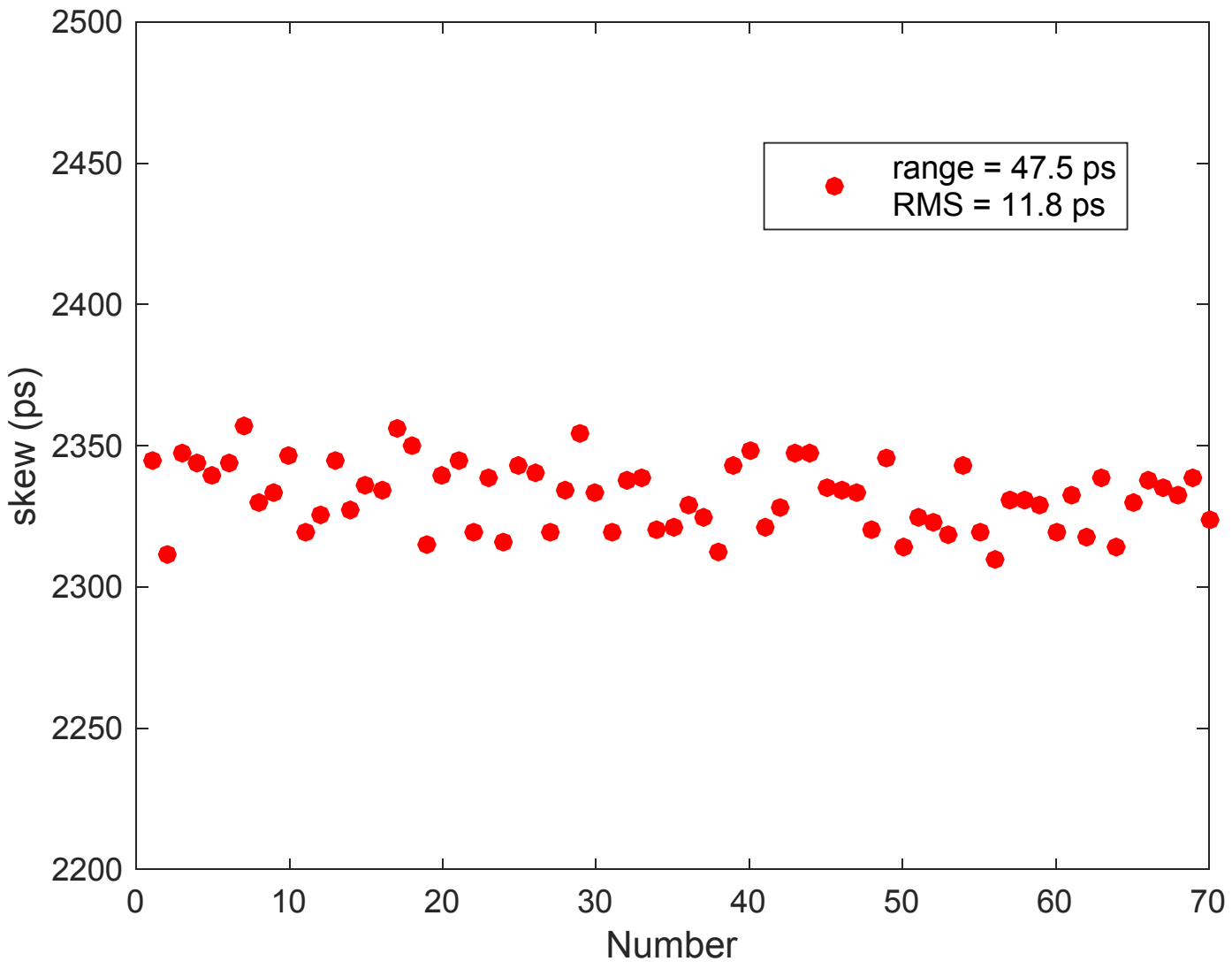

**Figure 12.** Result of clock synchronization test.

We also conducted tests to evaluate the performance with varying ambient temperature. As shown in figure 13, we placed one prototype module and the corresponding 400-meter fiber in a climate chamber while the other branches are placed in a constant temperature environment. We implemented FPGA-based Time-to-Digital Converters (TDCs) [16] in these two modules, and used a common signal as the inputs of the TDCs. And then we can obtain the compensation precision by observing the fluctuation of time difference measurement results. We changed the temperature in the climate chamber from -10 °C to around 50 °C, and conducted tests. The time of the input signal is digitized by the FPGA TDC and transferred through WR switch to PC by the SiTCP.

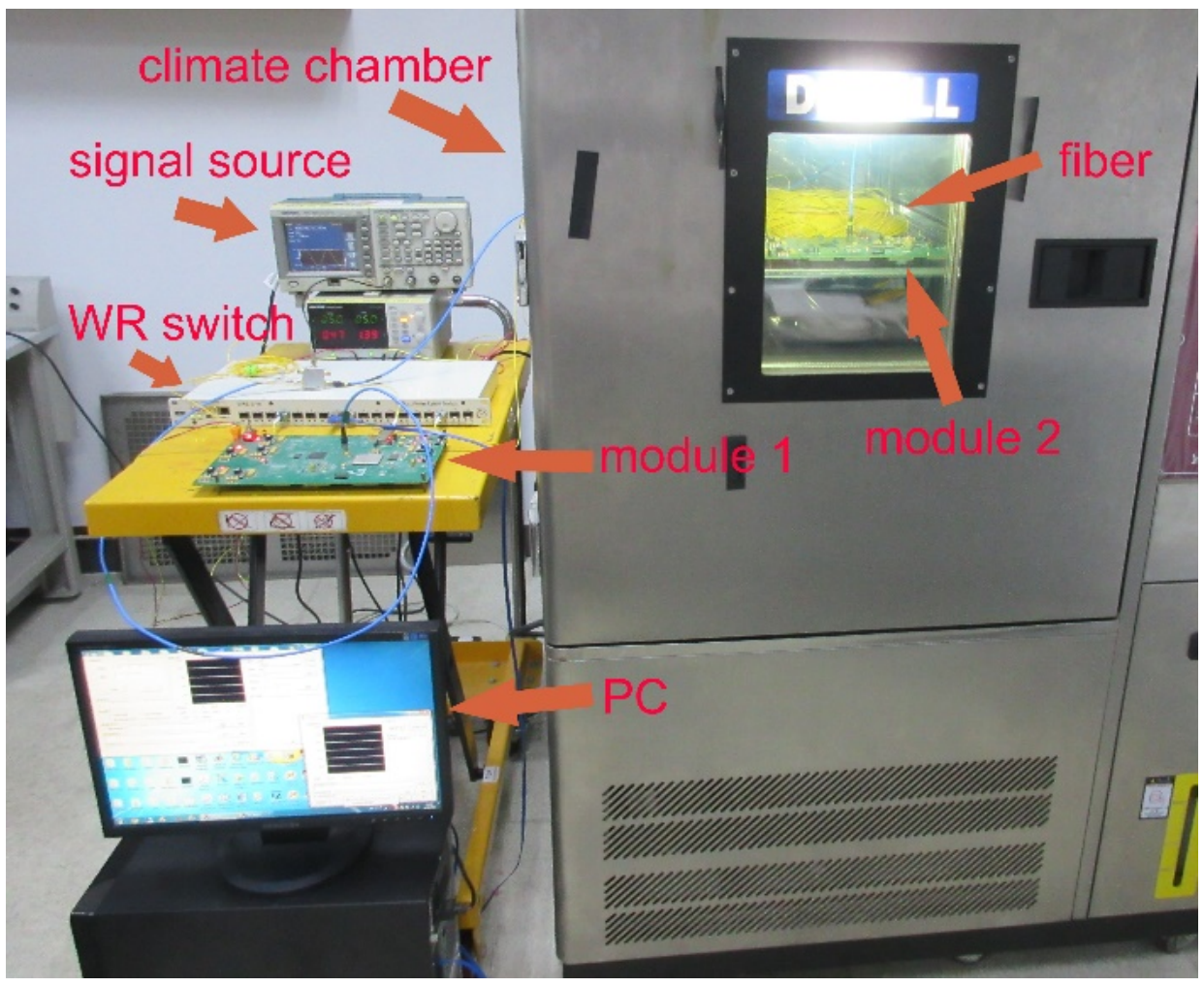

**Figure 13.** Temperature drift test in laboratory.

As shown in figure 14, a clock phase compensation precision better than 50 ps is achieved in a temperature range of around 60 °C.

.

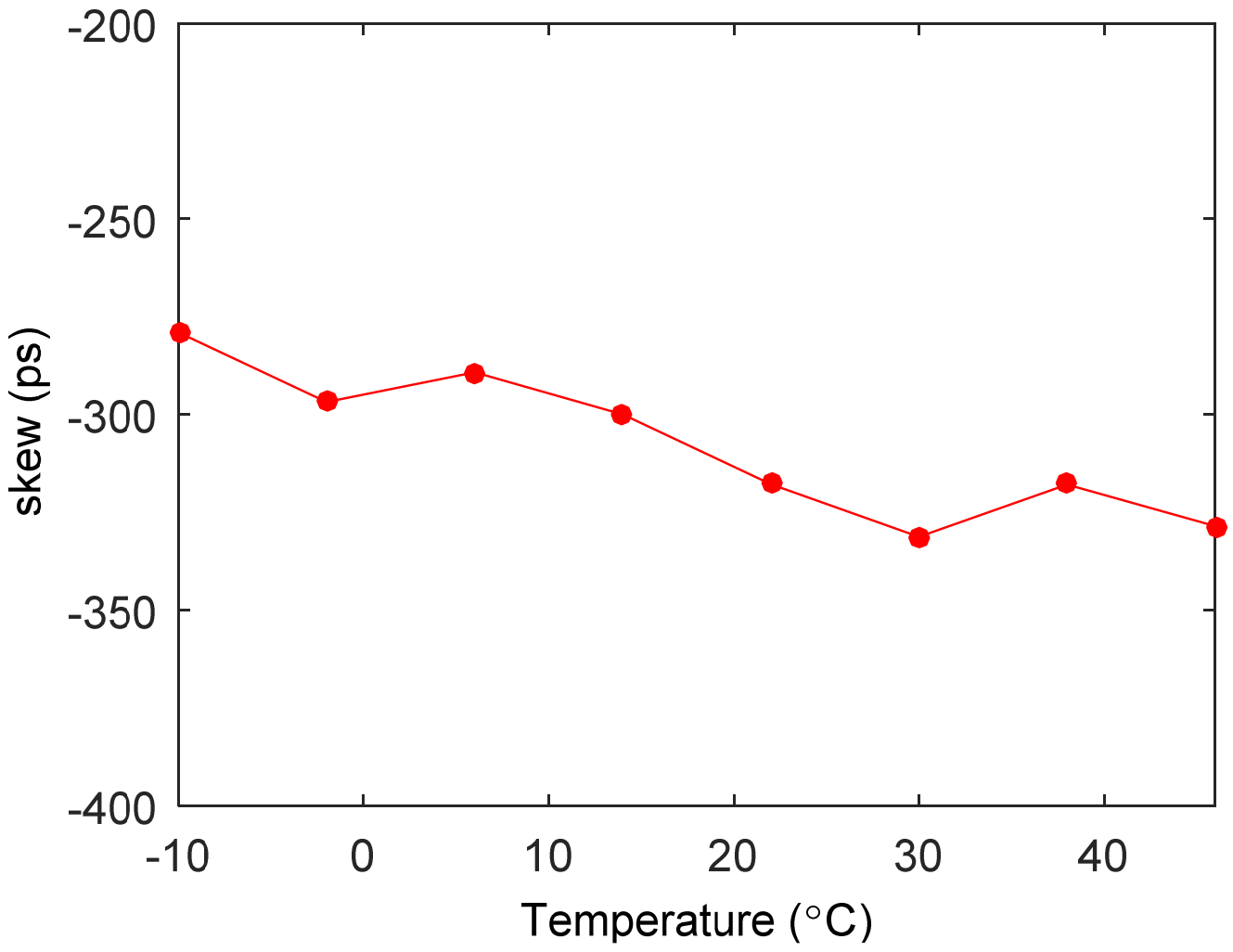

**Figure 14.** Test result of the compensation precision with varying temperature

### 3.2 Data Transfer Rate Test

We also conducted tests to evaluate the data transfer performance. Data from the prototype module are mixed with the clock signal and transmitted to the WR switch. The data are then extracted by the switch and further sent to a remote PC. Then we can calculate the data transfer rate on the PC, as shown in figure 15. The 24-hour test results indicate that the data transfer speed is around 400 Mbps, beyond the application requirements.

.

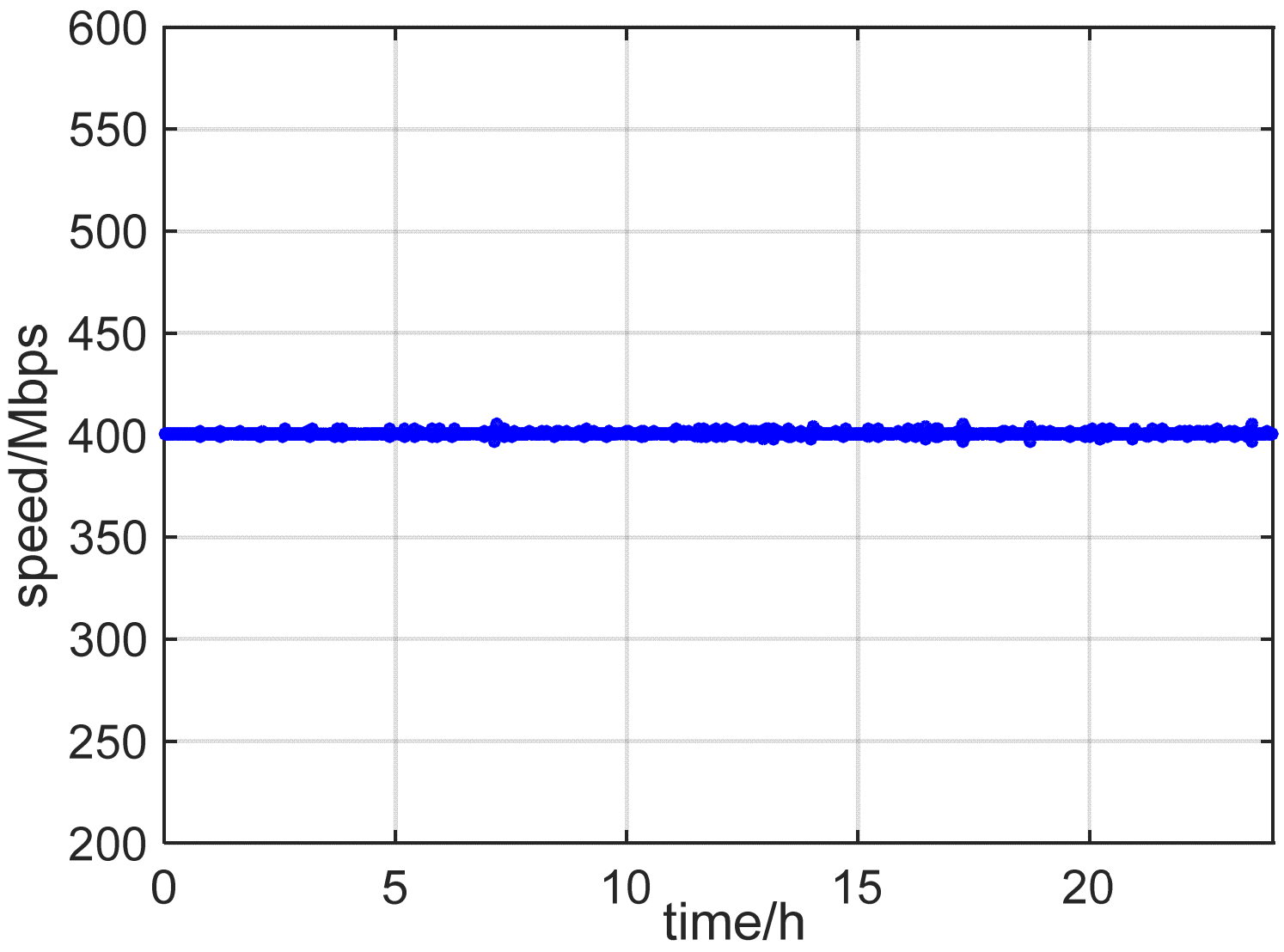

**Figure 15.** Test result of the TCP/IP data transmission

Considering that in actual application multiple FEEs will transfer data to one WR switch and further to DAQ, we also conducted tests to evaluate the performance in this situation. In this situation, the data rate is limited not only by the prototype modules but also by the performance of the WR switch. As shown in figure 16, we used four modules connected to one WR switch, and one PC connecting to another port of the switch. Figure 17 shows the test result. As for each module, a data speed of 180 Mbps is achieved, and a total data rate of 720 Mbps is achieved, which is beyond the readout requirements of the WCDA.

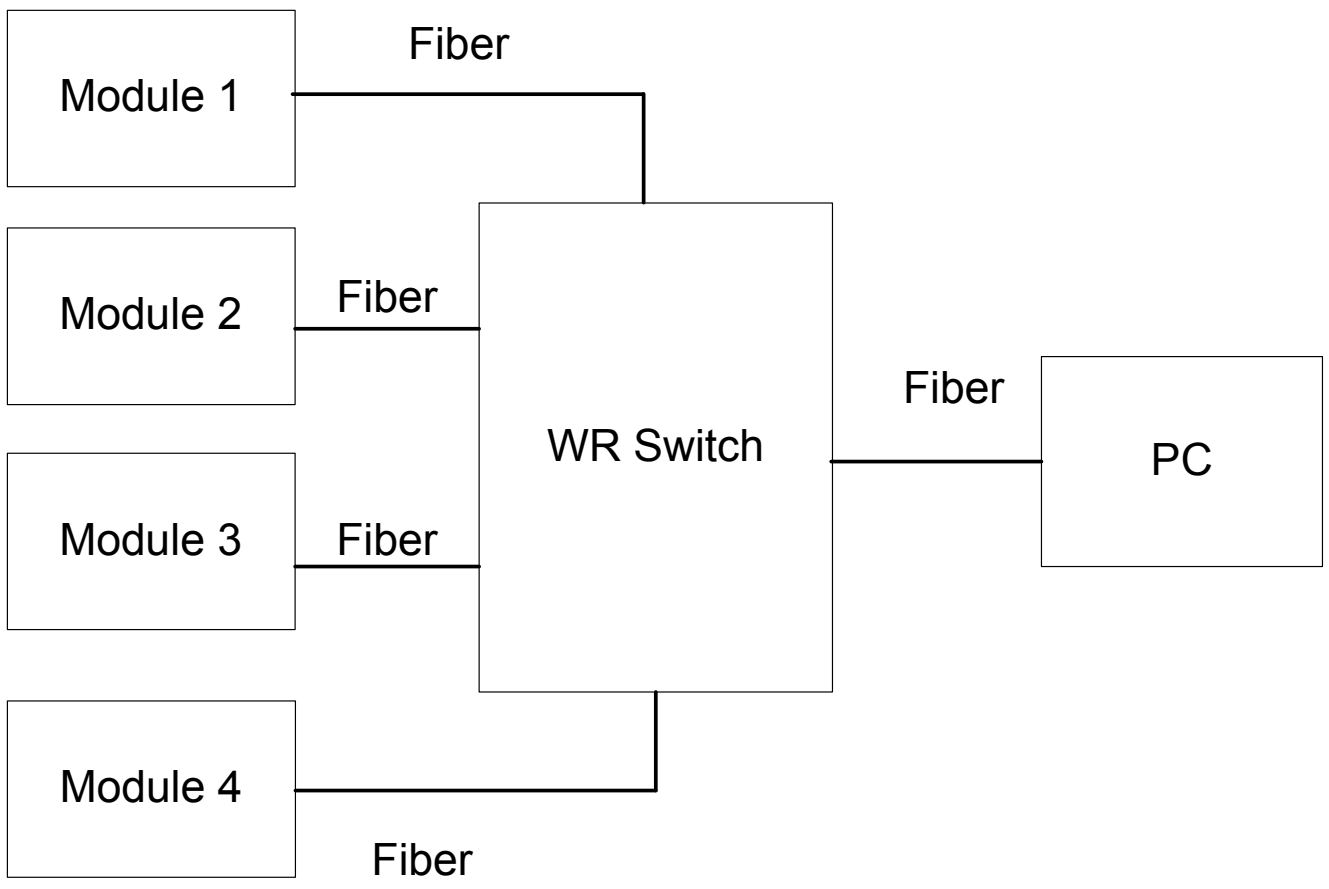

**Figure 16.** Test bench to evaluate the data rate of multiple FEEs with one WR switch

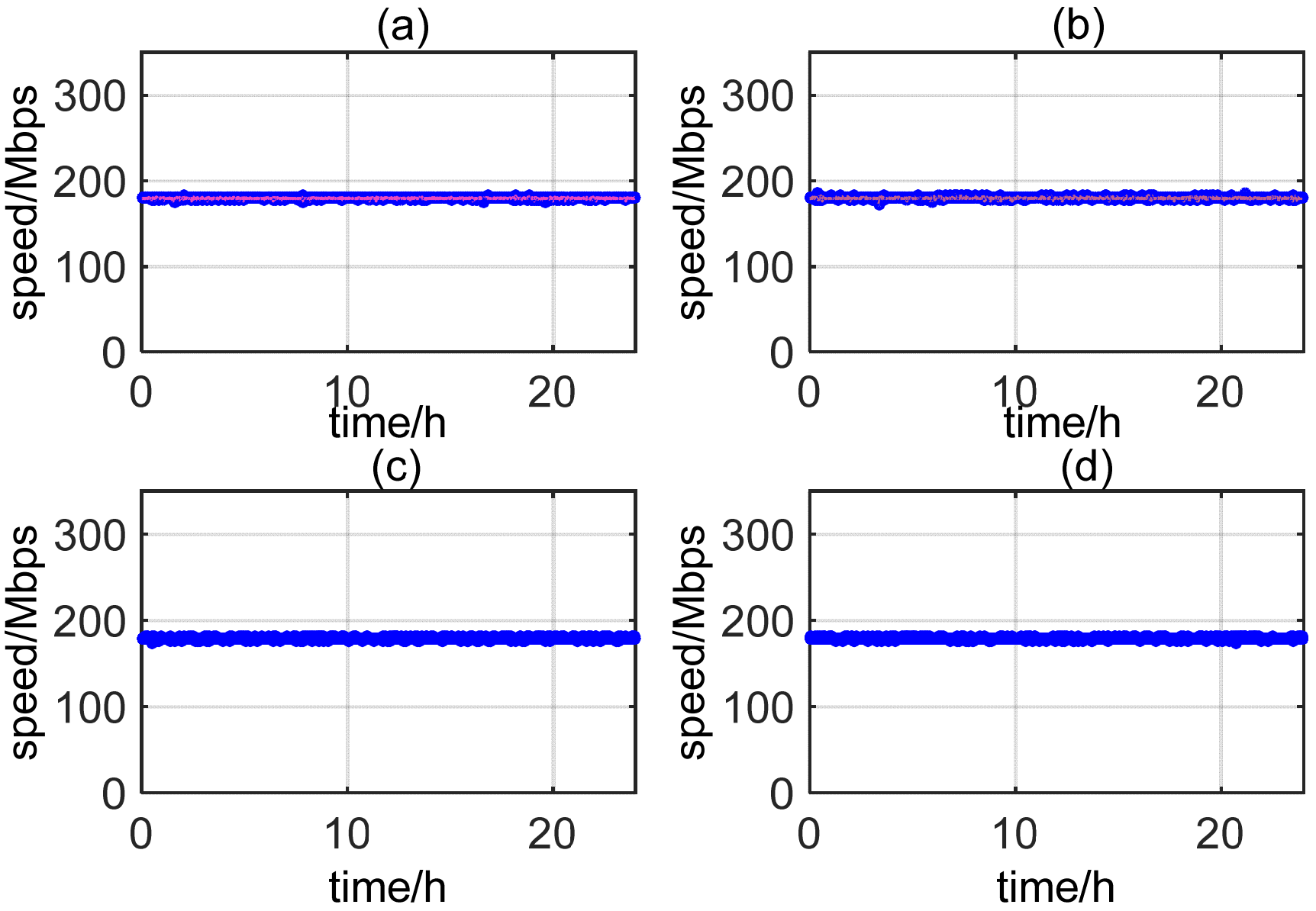

**Figure 17.** Data rate test result of multiple FEEs with one WR switch

### 3.3 Data Transfer Integrity Test

Based on the handshaking and retransmission mechanism, the TCP/IP protocol itself can guarantee good validity of the data. However, the data checking is based on CRC, which means that only part of the information participates in this checking process. Therefore, there still exists possibility of data errors with status of CRC error clean. Besides, the reliability of other parts in the data interface is not included in the CRC checking process of the TCP/IP protocol. To confirm the reliability of the overall data transfer interface that we designed, we conducted data transfer integrity tests referring to the standard Bit Error Rate test method. As shown in figure 18, a pseudo-random bit sequence (PRBS-7) is generated by the PRBS-7 generator block in the FPGA, and then is transferred through the SiTCP to the PC. On the PC, we designed a program in C language based on the same structure of the pseudo-random bit generator. With the same initial value, the output of these two generators should be same. By comparing the data received from SiTCP and the local pseudo-random bit stream, we can calculate the error bit rate. As shown in Table.1, the results indicate that the error bit rate is below $10^{-13}$, which is good enough for our application.

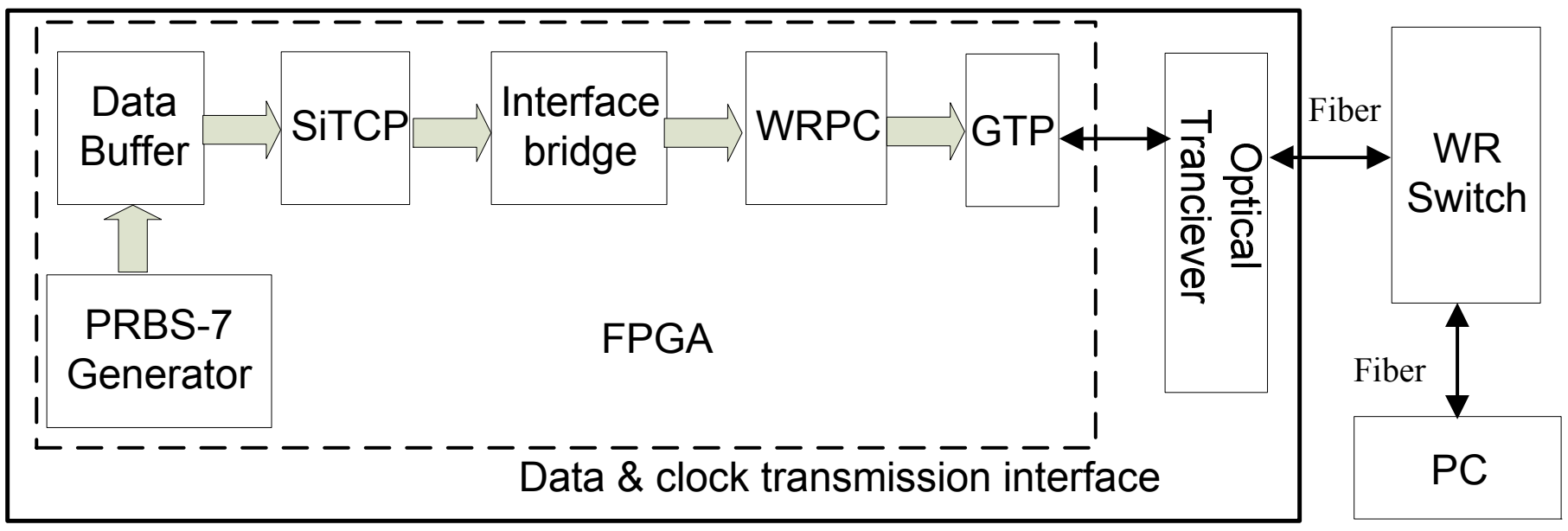

**Figure 18.** Test bench to evaluate the data integrity

**Table 1.** The result of the error bit rate test.

| Data transfer (bit) | Test time (h) | Error bit | Error bit rate |
|---|---|---|---|
| $1.38\times10^{13}$ | 24 | 0 | $<1\times10^{-13}$ |

## 4. Conclusion

To address the issues of high quality clock signal distribution and synchronization, as well as high-speed TCP/IP based data transfer due to “triggerless” architecture, we designed the clock and data transfer interface for the WCDA in LHAASO. This interface is also required to be compatible with the WR principle employed in the whole LHAASO project, and meanwhile, commands, data, and clock signal need to be mixed together and transmitted through the same fiber. Key techniques are discussed in this paper. We also designed prototype modules and conducted tests to evaluate the performance. Test results indicate that the clock synchronization precision is better than 50 ps over a 400 meters distance while the TCP data transfer speed is up to around 400 Mbps for each module, and 720 Mbps for four modules sharing one up link port, which are all beyond the application requirements.

## Acknowledgments

This work is supported by Knowledge Innovation Program of the Chinese Academy of Sciences (KJCX2-YW-N27), National Natural Science Foundation of China (11175174) and the CAS Center for Excellence in Particle Physics (CCEPP).